\documentclass[aps,pre,twocolumn,superscriptaddress,showpacs]{revtex4}
\usepackage{amsfonts,amssymb,amsmath,latexsym,epsfig,times} 
\begin{document}

\title{Enhancing complex-network synchronization}

\author{Adilson E. Motter}
\email{motter@mpipks-dresden.mpg.de}
\affiliation{Max Planck Institute for the Physics of Complex Systems,
N\"othnitzer Strasse 38, 01187 Dresden, Germany}

\author{Changsong Zhou}
\email{cszhou@agnld.uni-potsdam.de}
\affiliation{Institute of Physics, University of Potsdam
PF 601553, 14415 Potsdam,  Germany}

\author{J\"{u}rgen Kurths}
\affiliation{Institute of Physics, University of Potsdam
PF 601553, 14415 Potsdam,  Germany}


\begin{abstract}
Heterogeneity in the degree (connectivity) distribution has been shown to
suppress synchronization in networks of symmetrically coupled oscillators with
uniform coupling strength (unweighted coupling).  Here we uncover a condition for enhanced
synchronization in directed networks with weighted coupling.  We show that, in
the optimum regime, synchronizability is solely determined by the average degree
and does not depend on the system size and the details of the degree
distribution.  In scale-free networks, 
where the average degree may increase with heterogeneity,
synchronizability is drastically enhanced
and may become positively correlated with heterogeneity, 
while the overall cost involved in the network
coupling is significantly reduced as compared to the case of unweighted coupling.
\end{abstract}


\maketitle

Networks of dynamical elements serve as natural models for a variety of systems,
with examples ranging from cell biology to epidemiology to the Internet
\cite{reviews}.  Many of these complex networks display common structural
features, such as the small-world \cite{sw} and scale-free properties \cite{sf}.
Small-world networks (SWNs) exhibit short average distance between nodes and
high clustering \cite{sw}, while scale-free networks (SFNs) are characterized by
an algebraic, highly heterogeneous distribution of degrees (number of links per
node) \cite{sf}.  The interplay between structure and dynamics has
attracted a great deal of attention, especially in connection with the problem
of synchronization of coupled oscillators
\cite{watts:book,sync,BP:2002,NMLH:2003}.  The ability of a network to
synchronize is generally enhanced in both SWNs and random SFNs as compared to
regular lattices \cite{watts:book,sync,BP:2002,NMLH:2003,GDLP:2000}.  This
enhancement was previously believed to be due to the decrease of the average
distance between oscillators.  Recently, it was shown that random networks with
strong heterogeneity in the degree distribution, such as SFNs, are much more
difficult to synchronize than random homogeneous networks \cite{NMLH:2003}, even
though the former display smaller average path length \cite{CH:2003}.
This suggests that, although structurally advantageous \cite{AJB:2000},
the scale-free property may be dynamically detrimental.
Here we present a solution to this problem.

A basic assumption of previous work is that the oscillators are coupled
symmetrically and with the same coupling strength.  Under the
assumption of symmetric coupling, the maximum synchronizability is indeed
achieved when the coupling strength is uniform \cite{Wu:2003}.  But to get a
better synchronizability the couplings are not necessarily symmetrical.  Many
realistic networks are actually directed \cite{reviews} and weighted \cite{w}.
In particular, the communication capacity of a node is likely to saturate when
the degree becomes large.

In this Letter, we study
the impact that asymmetry and saturation of
connection strength have on the synchronization dynamics on complex networks.
As a prime example, we consider complete synchronization of linearly coupled
identical oscillators, namely
\begin{equation}
\dot{x_i}=f(x_i)-\sigma\sum_{j=1}^{N}G_{ij}h(x_j), \;\;\; i=1,\ldots N,
\label{eq1}
\end{equation}
where $f=f(x)$ describes the dynamics of each individual oscillator, $h=h(x)$ is
the output function, $G=(G_{ij})$ is the coupling matrix, and $\sigma$ is the
overall coupling strength.
In the case of symmetrically coupled oscillators with uniform coupling strength,
$G$ is the usual (symmetric) Laplacian matrix $L=(L_{ij})$  \cite{laplacian}.
For $G_{ij}=L_{ij}$,  heterogeneity in the degree distribution suppresses
synchronization \cite{NMLH:2003}.
In order to enhance the synchronizability of heterogeneous networks,
we propose to scale the coupling strength by the degrees
of the nodes.  For specificity, we consider
\begin{equation}
G_{ij}=L_{ij}/k_i^{\beta},
\label{eq2}
\end{equation}
where  $k_i$ is the degree of node $i$ and  $\beta$ is a tunable parameter.
The underlying network associated with the Laplacian matrix $L$ 
is undirected and unweighted, but with the introduction of the weights in
eq. (\ref{eq2}), the network of couplings
becomes not only weighted but also directed because the resulting matrix $G$ is in
general asymmetric.
We say that the network or coupling is weighted when $\beta\neq 0$  and
unweighted when $\beta=0$.

The variational equations governing the linear stability of a synchronized state
$\{ x_i(t)=s(t), \forall i\}$ can be diagonalized into $N$ blocks of the form
$\dot{\eta}=\left[ Df(s) - \alpha Dh(s)\right]\eta$, where
$\alpha=\sigma\lambda_i$, and $\lambda_i$ are the eigenvalues of the coupling
matrix $G$, ordered as $0=\lambda_1\le \lambda_2\cdots \le\lambda_N$ (see below).  The
largest Lyapunov exponent $\Lambda(\alpha)$ of this equation can be regarded as a
master stability function, which determines the linear stability of the
synchronized state \cite{msf}:  the synchronized state is stable if
$\Lambda(\sigma\lambda_i)<0$ for $i=2,\ldots N$.  (The eigenvalue $\lambda_1$
corresponds to a mode parallel to the synchronization manifold.)  For many
widely studied oscillatory systems \cite{BP:2002,msf}, the master stability
function $\Lambda(\alpha)$ is negative in a finite interval
$(\alpha_1,\alpha_2)$.  Therefore, the network is synchronizable for some
$\sigma$ when the eigenratio $R=\lambda_N/\lambda_2$ satisfies $R <\alpha_2/\alpha_1$.
The ratio $\alpha_2/\alpha_1$ depends only on the
dynamics ($f$, $h$, and $s$), while the eigenratio $R$ depends only on the
coupling matrix $G$.  The problem of synchronization is then reduced to the
analysis of eigenvalues of the coupling matrix \cite{BP:2002}:  the smaller the
eigenratio $R$ the more synchronizable the network.

Here we show that, as a function of $\beta$, the eigenratio $R$ has a global
minimum at $\beta=1$.  In large networks with some degree of randomness, the
eigenratio at $\beta=1$ is primarily determined by the average degree $k$ of the
network and {\it does not} depend on the degree distribution and system size, in
sharp contrast with the case of unweighted coupling ($\beta=0$), where
synchronization is strongly suppressed as the heterogeneity or number of
oscillators is increased.  Furthermore, we show that the total cost involved in
the network coupling is significantly reduced for $\beta=1$ when compared to
$\beta=0$.  As a result, structural robustness \cite{AJB:2000}
and improved synchronizability can coexist in scale-free and
other heterogeneous networks.

In matrix notation, eq.~(\ref{eq2}) can be written as $ G=D^{-\beta}L$, where
$D=$ diag$\{k_1,k_2,\ldots k_N\}$ is the diagonal matrix of degrees.  From the
identity $\det (D^{-\beta}L-\lambda I)=\det (D^{-\beta/2}LD^{-\beta/2}-\lambda
I)$, valid for any $\lambda$, we have that the spectrum of eigenvalues of matrix
$G$ is equal to the spectrum of a symmetric matrix defined as
$H=D^{-\beta/2}LD^{-\beta/2}$.  As a result, all the eigenvalues of matrix $G$
are real.
Moreover, because $H$ is positive semidefinite,  all the eigenvalues
are nonnegative and, because the rows of $G$ have zero sum, the
smallest eigenvalue $\lambda_1$ is always zero, as assumed above.  
If the network is connected, $\lambda_2>0$ 
for any finite $\beta$.
For spectral properties 
of unweighted networks, see refs.~\cite{chung:book,other_ref,CLV:2003,DGMS:2003}.

\begin{figure}[pt]
\begin{center}
\epsfig{figure=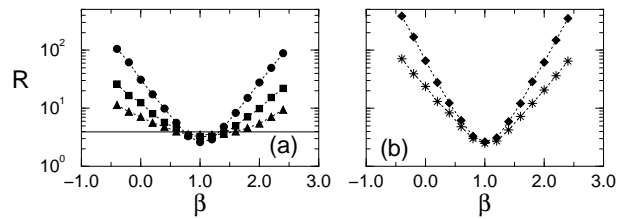,width=8.0cm}
\caption{
Eigenratio $R$ as a function of $\beta$: 
(a) random SFNs with $\gamma=3$
($\bullet$), $\gamma=5$ (${\scriptscriptstyle \blacksquare}$), $\gamma=7$
(${\scriptstyle \blacktriangle}$), and 
$\gamma=\infty$ (solid line), for $k_{min}=10$; (b) networks with expected
scale-free sequence (${\scriptstyle \blacklozenge}$) for $\gamma=3$ and ${\tilde
k}_{min}=10$, and growing SFNs (${\scriptstyle \bigstar}$) for $\gamma=3$ and
$m=10$.  Each curve is the result of an average over 50 realizations for
$N=1024$.
}
\label{fig1}
\end{center}
\end{figure}

We first examine the dependence on $\beta$.  Physically, we expect the
synchronizability to be strongly influenced by the strength of the input
coupling at each oscillator.  When $\beta < 1$, oscillators with larger degree
are more strongly coupled than oscillators with smaller degree.  When $\beta > 1$, the opposite happens.  
Because $(\alpha_1,\alpha_2)$ is finite,
for the network to synchronize, the overall coupling
strength $\sigma$ must be large enough to synchronize the least coupled
oscillators and small enough to synchronize the most coupled ones
(i.e., the synchronizability of these oscillators is expected to be
primarily determined by the modes associated with the eigenvalues $\lambda_2$ and $\lambda_N$, respectively). 
Therefore, for both $\beta < 1$ and $\beta > 1$,
some oscillators are more strongly coupled than others, and the ability of the
network to synchronize is limited by those oscillators that are least and most
strongly coupled.  We then expect the network to achieve maximum
synchronizability at $\beta=1$.  In fig.~\ref{fig1}, we show the numerical
verification of this hypothesis on 
three different models of SFNs, defined as follows:
\begin{description}
\item {(A)} {\it Random SFNs} \cite{NSW:2001} --- Each node is assigned to have
a number $k_i $ of ``half-links'' according to the probability distribution $P(k)\sim
k^{-\gamma}$, where $\gamma$ is a scaling exponent and $k\ge k_{min}$.  The
network is generated by randomly connecting these half-links to form links,
prohibiting self- and repeated links.  In the limit $\gamma=\infty$,
all the nodes have the same degree $k= k_{min}$.
\item {(B)} {\it Networks with expected scale-free sequence } \cite{CLV:2003}
--- The network is generated from a sequence $\tilde{k}_1, \tilde{k}_2, \ldots
\tilde{k}_N$,  where $\max_i\tilde{k}_i^2<\sum_i \tilde{k}_i$, so that links are
independently assigned to each pair of nodes $(i,j)$ with probability
$p_{ij}=\tilde{k}_i\tilde{k}_j/\sum_i \tilde{k}_i$.  When the expected degrees
$\tilde{k}_i\ge \tilde{k}_{min}$ follow the distribution $P(\tilde{k})\sim
\tilde{k}^{-\gamma}$, we have a network with expected scale-free sequence.
\item {(C)} {\it Growing SFNs} \cite{LLYD:2002} --- We start with a fully
connected network with $m$ nodes and at each time step  
a new node with $m$ links is added to the network.  
Each new link is connected to a node $i$ in the network with probability $\Pi_i\sim
(1-p) k_i + p$, where $0\le p \le 1$ is a tunable parameter.  
For large degrees,
the scaling exponent of the resulting network is $\gamma=3+p[m(1-p)]^{-1}$.
For $p=0$, we recover the Barab\'asi-Albert model \cite{sf}.  
\end{description}
\noindent As shown in fig.~\ref{fig1}, a pronounced minimum for the eigenratio
$R$ at $\beta=1$ is observed in each case.  
A similar 
minimum for $R$ at $\beta=1$ is also observed in many
other models of complex networks, including the Watts-Strogatz model \cite{sw} of SWNs.
The only exception is the class of homogeneous networks,
where all the nodes have the same degree $k$.  In this case, the weights can be
factored out and $R$ is independent of $\beta$, as shown in fig.~\ref{fig1}(a)
for random 
homogeneous networks
with $k=10$ (solid line).

\begin{figure}[pt]
\begin{center}
\epsfig{figure=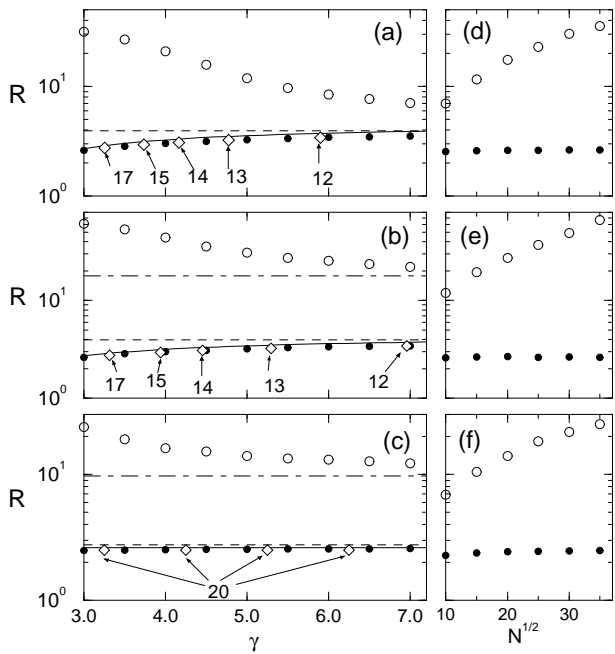,width=8.0cm}
\caption{
(a-c) Eigenratio $R$ as a function of the scaling exponent $\gamma$:  (a) random
 SFNs, (b) networks with expected scale-free sequence, and (c) growing SFNs, for
 $\beta=1$ ($\bullet$) and $\beta=0\;$ ($\circ$).  The other parameters are the
 same as in fig.~\ref{fig1}.  Also plotted are the bound of eq.  (\ref{10})
 (solid line) and $R$ at $\gamma=\infty$ for $\beta=1$ (dashed line) and
 $\beta=0$ (dot-dashed line).  The ${\scriptstyle \lozenge}$ symbols correspond
 to random homogeneous networks with the same average degree of the
 corresponding SFNs, as indicated in the figure.  (d-f) $R$
 as a function of the system size for $\gamma=3$ and the models in (a-c),
 respectively.  The legend is the same as in (a-c).
}
\label{fig2}
\end{center}
\end{figure}

In heterogeneous networks, the synchronizability is significantly enhanced when
the coupling is suitably weighted, as shown in fig.~\ref{fig2} for SFNs with
$\beta=1$.  In SFNs, the heterogeneity (variance) of the degree distribution
increases as the scaling exponent $\gamma$ is reduced.  When the coupling is
unweighted ($\beta=0$), the eigenratio $R$ increases with heterogeneity, but the
eigenratio does not increase and may even decrease with heterogeneity when the
coupling is weighted ($\beta=1$), as shown in figs.~\ref{fig2}(a-c).  The
enhancement is particularly large for small $\gamma$, where the networks are
highly heterogeneous (note the logarithmic scale in fig.~\ref{fig2}).  The
networks become more homogeneous as $\gamma$ is increased.  In the limit
$\gamma=\infty$, random SFNs converge to random homogeneous networks with the
same degree $k_{min}$ for all nodes [fig.~\ref{fig2}(a)], while networks with
expected scale-free sequence converge to Erd{\H o}s-R\'enyi random networks
\cite{ballobas:book}, 
which have links assigned with the same probability
between each pair of nodes [fig.~\ref{fig2}(b)].
As one
can see from fig.~\ref{fig2}(b), the synchronizability is strongly enhanced even
in the relatively homogeneous Erd{\H o}s-R\'enyi model; such an enhancement
occurs also
in growing networks [fig.~\ref{fig2}(c)].  
Surprisingly, for $\beta=1$, the eigenratio $R$ turns out
to be well approximated by the corresponding eigenratio of random homogeneous
networks with the same average degree [figs.~\ref{fig2}(a-c)].  Moreover, the
eigenratio $R$ appears to be independent of the system size for $\beta=1$ in
large SFNs, in contrast to the unweighted case, where $R$ increases strongly with
the number of oscillators [figs. \ref{fig2}(d-f)].

We now present an approximation for the eigenratio $R$ that supports and
extends our numerical observations.
In what follows we focus on the case $\beta=1$.
Based on results of ref.~\cite{CLV:2003} for random networks
with arbitrary expected degrees,
which includes important SFNs, 
we get 
\begin{equation}
\max \{1-\lambda_2,\lambda_N-1\}=
[1+o(1)]\frac{2}{\sqrt{\tilde{k}}},
\label{9}
\end{equation}
where $\tilde{k}$ is the average expected degree. 
This
result is rigorous for networks with a given expected degree 
sequence $\tilde{k}_1, \tilde{k}_2,\ldots\tilde{k}_N$, as defined in the model (B) above.
The assumption for this result is
$\tilde{k}_{min}\equiv\min_i\tilde{k}_i$ to be large as compared to
$\sqrt{\tilde{k}}\ln^3N$, but our numerical simulations suggest that this
assumption can be released considerably because eq.~(\ref{9}) is observed to
hold for  $\tilde{k}_{min}$ as small as $2\sqrt{\tilde{k}}$. 
Having  released this assumption,
from eq.~(\ref{9}) we have the following explicit upper bound for the
eigenratio in large networks:
\begin{equation}
R\le  \frac{1+2/\sqrt{\tilde{k}}}{1-2/\sqrt{\tilde{k}}}.
\label{10}
\end{equation}
Therefore, the eigenratio is bounded by a function of the average degree, 
which does not depend on the system size, in agreement with the
results in figs.~\ref{fig2}(d-f).  
Moreover,
we expect $R$ to approach the upper bound in eq.~(\ref{10}) because the semicircle law holds and the spectrum is symmetric
around 1 for $\tilde{k}_{min}\gg \sqrt{\tilde{k}}$ in the thermodynamical limit \cite{CLV:2003,DGMS:2003}.  
This prediction is confirmed numerically  under much weaker conditions, as
shown in figs.~\ref{fig2}(a-c), where one can see a remarkable agreement 
between the approximate and exact values of $R$ for all three models of SFNs.
Since the bound in
eq.~(\ref{10}) does not depend on the degree distribution, this result
also explains the agreement between the eigenratio for weighted SFNs
[figs.~\ref{fig2}(a-c), $\bullet$] and the eigenratio for  
random homogeneous networks
with the same average degree [figs.~\ref{fig2}(a-c), ${\scriptstyle \lozenge}$].  
A similar agreement is observed in many other complex networks.

\begin{figure}[pt]
\begin{center}
\epsfig{figure=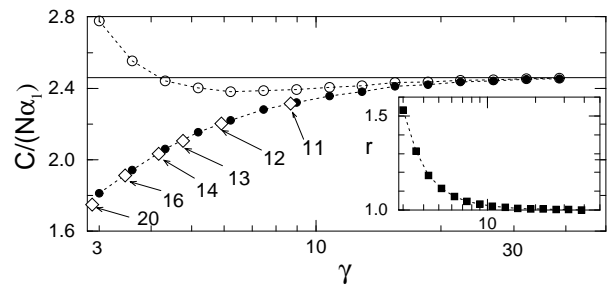,width=8.0cm}
\caption{Normalized cost as a function of the scaling exponent $\gamma$ for
random SFNs with $\beta=1$ ($\bullet$) and $\beta=0$ ($\circ$),
and for random homogeneous networks with the same average degree (${\scriptstyle
\lozenge}$).  The solid line corresponds to $\gamma=\infty$.
Inset:  Ratio $r=C_0/C_1$ of the cost for $\beta=0$ ($C_0$) and
$\beta=1$ ($C_1$) as a function of $\gamma$.  The other parameters are the same as in
fig.~\ref{fig1}.}
\label{fig3}
\end{center}
\end{figure}

It is also important to study the influence of  degree correlation and clustering
on the synchronizability of the networks.
Our extensive
numerical computation on the models of refs.~\cite{ST:2001,N:2003} shows that
the eigenratio $R$ generally increases with increasing clustering and
assortativity in correlated networks.  However, a pronounced
global minimum for $R$ as a function of $\beta$ is always observed at $\beta=1$.
In addition, weighted networks at $\beta=1$ are  much more insensitive to
the effects of correlation than their unweighted counterparts.  The same tendency 
is observed in the growing model with aging of ref.~\cite{DM:2000}, which has
nontrivial clustering and correlation.  All together, these suggest that our
results are quite robust and expected to hold on real-world networks as well.

Now we address the important problem of the {\it cost} involved in the
connections of the network.  The cost $C$ is naturally defined as the total
strength of all the directed links, i.e., $C=\sigma_{min}\sum_{i}k_i^{1-\beta}$,
where $\sigma_{min}=\alpha_1/\lambda_2$ is the minimum overall coupling strength
for the network to synchronize.  Strikingly, in heterogeneous networks, the cost
for $\beta=1$ is considerably smaller than the cost for $\beta=0$
[fig.~\ref{fig3}].  The cost for $\beta=1$ is well approximated by the cost for
random homogeneous networks with the same average degree $k$, as indicated in
fig.~\ref{fig3}.  In this case, from above we have
$C/(N\alpha_1)=1/\lambda_2\approx 1/(1 - 2/\sqrt{k})$, and the cost is reduced
as $k$ is increased, approaching $C/(N\alpha_1)=1$ for large globally coupled
networks.  Therefore, cost reduction is another important advantage of the
weighted coupling.

In summary, we have introduced a model of directed networks with weighted
coupling which, we believe, can serve as a paradigm to address various issues
regarding dynamics on complex networks.  Within this model, we have shown that
suitably weighted networks display significantly improved synchronizability and
lower cost.  An important implication of our findings is that weighted SFNs can
exhibit enhanced complete synchronization.  We expect our results to be relevant
for both network design and the understanding of dynamics in natural systems,
such as synchronization in neural networks \cite{neural} and synchronization of
epidemic outbreaks in networks of cities \cite{city}, where weighted couplings
are expected to play an important role \cite{bernd}.

\acknowledgments
A.~E.~M.~is supported by MPIPKS. C.~S.~Z.~and J.~K.~are  supported by SFB 555.


\begin{thebibliography}{0}
\bibitem{GDLP:2000}
A different behavior has been observed in networks of pulse oscillators~\cite{GDLP}.
\bibitem{laplacian}
The diagonal entries are $L_{ii}=k_i$.
The off-diagonal entries are $L_{ij}=-1$, if nodes $i$ and $j$ are connected, and  $0$ otherwise.
\end{thebibliography}
\end{document}